\documentclass[twocolumn]{revtex4}
\usepackage{graphics}

\begin{document}

\title{Effective theory of high-temperature superconductors}

\author{Igor F. Herbut}

\address{Department of Physics, Simon Fraser University, 
Burnaby, British Columbia, Canada V5A 1S6 }

\begin{abstract}
The field theory of a fluctuating
d-wave superconductor is constructed and proposed as an effective
description of superconducting cuprates at low energies.
The theory is used to resolve a puzzle posed by recent experiments
on superfluid density in severely underdoped $YBa_2 Cu_3 O_{6+x}$.
In particular, the
overall temperature dependence of the superfluid density at low dopings
is argued to be described well by the strongly anisotropic weakly interacting
Bose gas, and thus approximately
linear in temperature with an almost doping-independent slope.
\end{abstract}
\maketitle

The superconducting state of underdoped high-temperature
superconductors is long known to be anomalous. Whereas the pseudogap
temperature $T^*$ is high,
and only increasing with underdoping,
the superconducting transition temperature $T_c$, together with the
zero-temperature superfluid density $\rho(0)$ at the same time
is continuously vanishing \cite{uemura}. This is in stark contrast to
 the predictions of the standard BCS theory,
in which of course $T_c$ and $T^*$ would be essentially identical,
and $\rho(0)$ proportional to the electron density, i. e.
$\rho(0)\sim 1-x$, where $x$ is the number of holes per lattice site
(doping). It has recently been argued \cite{paramekanti}, \cite{igor1},
 that when pairing in the d-wave channel is accompanied by 
strong repulsion the amplitude of the 
order parameter $|\Delta|$ and $\rho(0)$
may behave oppositely as half-filling ($x=0$) is approached,
with the divorce of $T^*$ and $T_c$ arising as a natural consequence.
Disordering of the d-wave superconductor (dSC)
by quantum fluctuations thus appears, at least in this respect,
to be similar to doping of holes into the Mott insulator (MI) as described 
by the effective gauge theories of the t-J model \cite{wen}. 

There exist, however, experimental results that appear to contradict 
the theories of the fluctuating dSC  \cite{balents}, \cite{franz}, 
\cite{igor2}, or the gauge theories of the t-J model \cite{wen}.
Whereas these approaches correctly yield
$\rho(0)\sim x$ as has been seen, they also generally imply  that
the `effective charge' of quasiparticles becomes proportional to
doping  \cite{ng}, and thus $d\rho(T)/dT|_{T\rightarrow 0} \sim x^2 $,
which has not. On the contrary, the experiments in strongly underdoped
single crystals \cite{broun} and thin films \cite{lemberger} of
$YBa_2Cu_3 O_{6+x}$ (YBCO) show the superfluid density to be
approximately linear in temperature over most of the temperature range,
with an almost doping-independent slope, and
with a higher power-law emerging near $T=0$.
The critical region, almost 10 Kelvins wide at optimal doping, appears also 
to have completely disappeared at low dopings. The doping-independent
slope becomes particularly troubling in light of the recent 
measurements of heat transport in high-purity single crystals
of YBCO \cite{sutherland1}, in which
gapless quasiparticles in strongly underdoped samples 
indeed appear decoupled from the external magnetic field.
This would agree with the theoretical prediction
that quasiparticles should gradually become `neutral' 
with underdoping, but at the same time makes the behavior   
of the superfluid density all the more puzzling. 

In this Letter I present the field theory of the
quantum-fluctuating d-wave superconductor (dSC).
The spin sector, which was the primary focus of the earlier studies
\cite{franz}, \cite{igor2}, \cite{igor3}, has been embedded into a  
more general framework describing both spin and charge degrees of freedom
 at finite dopings.
The theory describes nodal quasiparticles coupled to the phase  
fluctuations of the order parameter. Quantum
fluctuations should arise from the Coulomb interaction, which becomes 
increasingly detrimental to phase coherence near half-filling \cite{igor1}.
Such an effective theory should provide a correct description at
energies below the amplitude $|\Delta|$. From this starting
point two main results are further derived. First,
the general theory  is shown to be considerably simplified by being rewritten 
in terms of the dual variables. Remarkably, in this 
form it resembles the effective SU(2) gauge theory
of the t-J model \cite{wen}. Second, the transformed theory
facilitates a simple, and a physically transparent calculation of the
ab-plane superfluid density $\rho (T)$ in the relevant doping regime.
Although the
quasiparticle `charge' is found to indeed be gradually vanishing,
on the scale of $T_c$ the region of temperatures where the 
slope of $\rho(T)$ would become too strongly doping-dependent
becomes negligible at low dopings. In the leading approximation,
the overall form of $\rho (T)$ is given by the condensate of the
strongly anisotropic  weakly-interacting  three-dimensional Bose gas. This,
on the other hand, is shown to be $\Delta \rho(T)\sim  T\ln(T/t)$, with $t$ 
as the Josephson inter-layer coupling, and thus also 
approximately linear in temperature, but with the slope at most
logarithmically dependent on doping. This also explains
the absence of a discernible critical region and a
slight curvature of the data near $T=0$ (Fig. 1).
Finally, properties of the insulating phase at $x=0$ are briefly discussed.

Let me begin by postulating the action for the low-energy quasiparticles
of a two-dimensional (2D) phase-fluctuating dSC, 
$S=\int_0^{1/T} d\tau \int d^2 \vec{x} L$, with the Lagrangian density
$L=L_\Psi + L_\Phi$, and
\begin{eqnarray}
L_\Psi = \bar{\Psi}_1 [\gamma_0 (\partial_\tau -i a_0)
+ v_F \gamma_1 (\partial_x-i a_x) + \\ \nonumber
 v_\Delta \gamma_2 (\partial_y-i a_y)] \Psi_1 
+ (1\rightarrow 2, x\leftrightarrow y)+ iJ_\mu (v_\mu +A_\mu),
\end{eqnarray}
\begin{eqnarray}
L_\Phi = \frac{i}{\pi }\epsilon_{\mu\nu\rho} (a_\mu,v_\mu) \partial_\nu
(A^- _\rho, A^+ _\rho)^{T} + \frac{K_\mu }{2} (v_\mu +A_\mu )^2 \\ \nonumber
+ \frac{h}{\pi }\epsilon_{0\mu\nu}\partial_\mu A^+ _\nu
+ \frac{1}{2}
\sum_{n=1}^2 |(\partial_\mu - i (A^+ _\mu + (-)^n A^- _\mu))\Phi_n|^2 \\ \nonumber
+ \tilde{\alpha} \sum_{n=1}^2 |\Phi_n|^2 
+ \frac{\tilde{\beta}_1}{2} (\sum_{n=1}^2 |\Phi_n|^2)^2
+ \frac{\tilde{\beta}_2}{2} \sum_{n=1}^2 |\Phi_n|^4.
\end{eqnarray}
Two four-component fermionic fields $\Psi_{1,2}$ describe the gapless
spin-1/2 excitations near the two pairs of diagonally opposed nodes.
$v_F $ and $v_\Delta\sim |\Delta|$ are
the two characteristic velocities of the low-energy spectrum,
$\{\gamma_\mu,\gamma_\nu \} = 2\delta_{\mu \nu}$, and
$J_\mu$ and $\bar{\Psi}_n \gamma_\mu \Psi_n$, $\mu=0,1,2$ are the charge
and spin currents, respectively, as defined in \cite{igor2}.
Whereas the Fermi velocity $v_F$ should be approximately
independent of $x$, $v_\Delta$ may be assumed to be a decreasing function of
doping \cite{paramekanti}, \cite{igor1}. $a_\mu$ and $v_\mu$ are the Berry and 
the Doppler $U(1)$ fields that furnish the coupling of the quasiparticles
to the fluctuating phase of the order parameter \cite{franz}; 
the integration over the auxiliary gauge fields $A^{\pm} _\mu $ constrains
$\epsilon_{\mu\nu\rho}\partial_\nu (a_\rho, v_\rho ) =
\pi (J_{\Phi_1} - J_{\Phi_2}, J_{\Phi_1} + J_{\Phi_2} )_\mu$, where
$J_{\Phi_{1,2}}$ are the vortex current densities. $K_0$ and
$K_{1,2}=K \sim E_F \sim 1-x $ are the bare compressibility and the
bare superfluid density, respectively, which derive from the
integration over the high-energy fermions.
$A_\mu$ is the physical electromagnetic
vector potential, and $h$ the bare chemical potential. The parameter $
\tilde{\alpha}$ tunes quantum fluctuations, and $\tilde{\beta}_{1,2}>0$
describe the short-range
repulsive interactions between vortex loops. Terms that are irrelevant
at low-energies \cite{babak} have been omitted. 

At $h=0$ $L_\Phi$ represents the continuum limit
of the lattice theory discussed in \cite{igor2}. Its form may 
be also understood on the basis of symmetry:
1) the usual electromagnetic gauge invariance under
$A_\mu \rightarrow A_\mu +\partial_\mu \chi$, 
2) the internal gauge invariance under $a_\mu \rightarrow a_\mu +\partial_\mu
\chi$, 3) the Ising  symmetry under $\Phi_1\leftrightarrow \Phi_2$,
$a_\mu \leftrightarrow -a_\mu$, spin up$\leftrightarrow$down, 
and 4) the gauge invariance under $A_\mu ^{\pm}
\rightarrow A_\mu ^{\pm} + \partial_\mu \chi^{\pm} $, 
together with the requirement of analyticity in $\Phi_{1,2}$ dictate
the form of $L$ as unique to the lowest non-trivial order in
the fields and their derivatives. An important feature of the Lagrangian 
is the addition of the chemical potential $h$, which is necessary to allow for 
a finite doping. It is introduced  
by shifting $A_0\rightarrow A_0 + ih$, after which it has been absorbed 
into a redefined $v_0$, as $v_0 +ih \rightarrow v_0$. It then appears only 
as the fictitious external `magnetic field' in the $\tau$-direction in
$L_\Phi$. As will be discussed shortly, having $h\neq 0$ is crucial for
obtaining the correct charge dynamics of the fluctuating dSC. 

The form of $L$ may also be justified on phenomenological grounds,
since, as discussed below, it describes a novel MI-dSC transition of
possible relevance to cuprates. It is convenient, however, 
to derive the equivalent {\it dual}
form of $L_\Phi$, more amenable at $h\neq 0$, first.
Duality is most precisely established on a lattice,
 and for the `hard-spin' version of the complex fields. Using a fairly
  standard set of transformations \cite{dasgupta}
 it is easy to show that, modulo analytic terms,
 \begin{eqnarray}
 \int (\prod_{r} d\phi_r d\vec{A}_r)
 e^{\sum_r \frac{1}{T} \cos(\Delta \phi - 2 \vec{A}) -\frac{i}{\pi}
 \vec{a}\cdot \Delta\times\vec{A}} = \\ \nonumber
 \lim_{t\rightarrow 0} \int (\prod_{r} d\psi_r )
 e^{\sum_r \frac{1}{t} \cos(\Delta \psi - \vec{a})
 -\frac{T}{8 \pi ^2 } (\Delta\times\vec{a})^2 } ,
 \end{eqnarray}
 where $r$ labels the sites of the 2+1D quadratic lattice, and $\Delta$ and
 $\Delta\times$ are the lattice gradient and the curl. Taking the
 continuum limit and going into the `soft-spin' representation
 \cite{dasgupta} this implies that 
\begin{eqnarray}
L_\Phi = \frac{K_\mu }{2} (v_\mu +A_\mu )^2
+h\sum_{n=1}^2 b^* _n (\partial_0- i (v_0 +(-)^n a_0)) b_n \\ \nonumber 
+ \frac{1}{2}
\sum_{n=1}^2 |(\partial_\mu - i (v _\mu + (-)^n a_\mu ))b_n|^2 \\ \nonumber
+ (\alpha-\frac{h^2}{2}) \sum_{n=1}^2 |b_n|^2 
+ \frac{\beta_1}{2} (\sum_{n=1}^2 |b_n|^2)^2
+ \frac{\beta_2}{2} \sum_{n=1}^2 |b_n|^4. 
\end{eqnarray}
$L_\Phi$ may therefore be alternatively understood as
describing the coupling of the fields $a_\mu$ and $v_\mu$ to
{\it two non-relativistic} bosonic fields of unit electromagnetic
charge, which are dual to the original vortex fields $\Phi_{1,2}$.
It is interesting to note that, although presumably more general,  
Eq. 4 is similar to the conjectured effective theory of
the s-flux phase within the $SU(2)$ gauge theory of the t-J model
\cite{wen}. Somewhat similar connection between the
U(1) gauge theory of the t-J model and the dSC
with topologically trivial phase fluctuations \cite{balents}
has been noted in the past \cite{dhlee}, \cite{alexandre}.

Consider the {\it superconducting phase} described by $L$.
Assuming $\alpha>0$, $\beta_{1,2} >0$ and minimizing $L_\Phi$,
 for $h>\sqrt{2\alpha}$
one finds $|\langle b_1 \rangle|^2 = |\langle b_2 \rangle|^2 
=(h^2-h_c ^2)/2(2\beta_1 +\beta_2 )$. To the quadratic order
$L_\Phi$ then reduces as
\begin{equation}
L_\Phi \rightarrow \frac{K_\mu }{2} (v_\mu + A_{\mu})^2
+ \frac{\rho_b (T) }{2} (v_\mu ^2 + a_\mu ^2) - ih n_b v_0 ,
\end{equation}
where $\rho_b(T) $ and $n_b$ are the superfluid 
and the total density of the bosons, respectively. Since both
$U(1)$ fields are massive, nodal
quasiparticles interact only via short-range
interactions, and therefore represent well-defined low-energy excitations,
conducting heat, for example. At low energies 
$L =L_{sp} + L_{ch}$, where the spin part of the Lagrangian to the leading
order is $L_{sp}= L_\Psi -iJ_\mu (v_\mu + A_\mu)
+ (\rho_b /2) a_\mu ^2 $ \cite{igor3}, and
\begin{equation}
L_{ch} =  iJ_\mu (v + A)_\mu 
+ \frac{K_\mu }{2} (v + A)_\mu^2 +\frac{\rho_b (T)}{2} v_\mu ^2
- i h n_b v_0.
\end{equation}
Setting $A_0=0$ and integrating over $v_0$ one finds 
\begin{eqnarray}
L_{ch} = \frac{J_0 ^2}{2 (K_0 +\rho_b(T) )} -\frac{h n_b }{K_0
+\rho_b(T) } J_0+ \\ \nonumber
 i\vec{J}\cdot (\vec{v} + \vec{A}) 
+ \frac{K}{2} (\vec{v} + \vec{A})^2 +\frac{\rho_b (T) }{2} \vec{v}^2, 
\end{eqnarray}
where $\vec{X}=(X_1, X_2)$.
The first term describes a short-range repulsion between fermions
and as such is irrelevant at low energies. The second determines 
the renormalized chemical potential: $\mu(T) =-h n_b
/(K_0 +\rho_b(T) )$. Since $x\sim -\mu(0) $, one finds
$ n_b \sim x$. The rest of
$L_{ch}$ determines the superfluid density.
Integrating $\vec{v}$ generates the term 
$ z \vec{J}\cdot \vec{A}$ in $L_{ch}$, with the ratio 
 $z= \rho_b(T)/( K+\rho_b (T) ) $ as the Fermi liquid
`charge renormalization' parameter. The electromagnetic field 
gradually decouples from quasiparticles with underdoping,
in accord with the observed insensitivity of thermal conductivity to the 
magnetic field in underdoped YBCO \cite{sutherland1}.

The suppression of the quasiparticle charge,
and particularly its effect on the
superfluid density, may be alternatively understood by integrating
fermions in $L$ {\it before} $\vec{v}$.  At $T\neq 0$, this 
reduces $K$ in Eq. 7 as: $K\rightarrow K(T)=
K- (2\ln(2)/\pi) (v_F/v_\Delta) T +O(T^2)$.
Integrating then $\vec{v}$ yields the total superfluid density: 
\begin{equation}
\rho(T)^{-1} = K(T)^{-1} + \rho_b(T) ^{-1}.  
\end{equation}
The last result, also known as 
Ioffe-Larkin rule \cite{wen}, is easily seen to hold
both in the underdoped (large $K$), and in the overdoped regime (large
$\rho_b (0)$). At $T=0$, therefore,
$\rho(0)= K\rho_b (0) /(K+ \rho_b (0) )\sim n_b \sim x$
at low $x$, which agrees qualitatively 
with more microscopic calculations \cite{paramekanti}, \cite{igor1}, as well
as with experiment \cite{uemura}. At $T\neq 0$, however,
$\rho_b (0) - \rho_b (T) \propto  T^3$ in 2D \cite{case},
and the leading temperature dependence of $\rho(T)$ at low T comes from  
the first, quasiparticle, term in Eq. 8. Differentiation gives
 $d\rho(T)/dT=(\rho(0)/K)^2 dK(T)/dT\sim z ^2$ at low $T$. 
Such a strong, $\sim \rho^2 (0) $, dependence of the slope
with doping, however, appears to be clearly contradicting the experiment
\cite{broun}, which defines our main problem.

\begin{figure}[t]
{\centering\resizebox*{80mm}{!}{\includegraphics{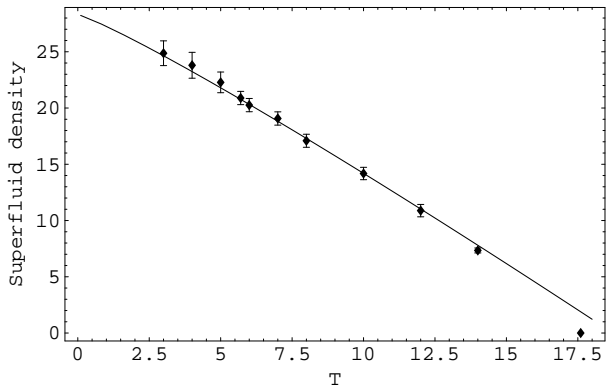}} \par}
\caption[]{ $H_{c1}$ in an underdoped YBCO \cite{broun}
converted to ab-plane superfluid density in Kelvins \cite{case} 
as $\rho= (4\pi H_{c1} / ((\ln (\kappa) +0.5 )20.7 Gauss)) (6.2 K)
(c/10 \mbox{\AA}) $, for $\kappa=100$ and the c-axis
anisotropy $10^{-4}$ \cite{david}. The line is $\rho_b (T)$ from
the Eq. 9. with $\rho_b (0)=28.3 K$. }
\label{sfdensity}
\end{figure}

While the reasoning from the previous paragraph is correct in principle,
the range of temperatures where it applies clearly depends on the specific
form of the bosonic term in Eq. 8. One needs therefore to  understand 
$\rho_b(T)$ in the underdoped region, which corresponds to the
{\it dilute} system of bosons, in
more detail. Before turning to the case of cuprates,
let me first briefly review a canonical example of bosons with a mass $m$
interacting with a two-body interaction $V(\vec{x})=\lambda\delta(\vec{x})$.
Dimensional analysis dictates that the superfluid density in an
isotropic 3D system may be written as
$\rho_b = r^{-3} G(m\lambda/r, mT r^2)$, where $r$ is the average distance
between particles, and $G(x,y)$ a dimensionless function of two
dimensionless arguments, with $G(0,0)=1$. The dilute limit
$r\rightarrow \infty$  is equivalent
therefore to the weakly interacting high-temperature limit, where 
$G \approx 1-(T/T_{BEC})^{3/2} $. This is just
the temperature dependence of the {\it condensate of non-interacting
bosons}, with $T_{BEC}$ being the Bose-Einstein condensation temperature.
At low temperatures, of course, interactions always turn
the temperature behavior into $\sim T^4$ (in 3D), but
this higher power-law sets in only below the characteristic
interaction temperature scale $\Delta T_\lambda \sim (m\lambda/r) T_{BEC}$.
Similarly, the critical region 
$ \Delta T_c \sim (m\lambda/r)^2 T_{BEC} $, so both $\Delta T_{\lambda,c}
\ll T_c \approx T_{BEC}$ in the dilute regime. This is all just another way
of phrasing the {\it irrelevancy} of short-range interactions near
the quantum critical point of 3D bosons.
Short-range interactions are irrelevant in 2D as well but only
marginally so, and consequently the weakly interacting regime is harder to
reach. Nevertheless, in either case the bosonic superfluid density 
 in the sufficiently dilute regime approaches the characteristic 
temperature dependence of the condensate, 
$\Delta \rho_b (T) \sim T^{D/2}$, except in the narrow
low-temperature and critical regions \cite{fisher}. 

What is the function $\rho_b(T)$ then in severely underdoped cuprates?
First, the phase coherence in the
superconducting state is known to be strongly anisotropic, but nevertheless
fully three-dimensional. We may include this feature by additionally coupling 
the different superconducting layers, each described by Eq. 4 and labeled
by $l$, via a weak Josephson coupling of the form
$t\sum_{l,n=1,2}b^* _{n,l} b_{n,l+1}$, with $t\ll \rho_b (0)$.
Second, in such a strongly
anisotropic system the Coulomb interaction
within a layer is screened well by the other layers, and except at
extremely low densities $\rho_b (0) \sim t$ it may indeed be taken to be
effectively short-ranged \cite{doniach}, as in Eq. 4.
Away from the ultimate quantum critical region of the theory (1)
where both the long-range nature of the Coulomb repulsion, 
and possibly the gauge field fluctuations, would finally become important,
such a short-range interaction is then irrelevant,  and
$\rho_b (T)$ to the zeroth order in interaction is just  
the condensate of the non-interacting system. In the anisotropic 3D
case this gives
$\rho_b (T) = \rho_b (0) - T F(t/T)/(2\pi)$, where
\begin{equation}
F(x) =  - \int_0 ^1 dz \ln( 1- e^{-x \sin^2 (\pi z/2)}),  
\end{equation}
with $F\approx 2.612 (T/\pi t)^{1/2}$ for $T\ll t$, and $F\approx \ln (T/t)$
for $t \ll T$. Since $t \ll \rho_b (0)$, 
on the scale of $T_c$ $\rho_b (T)$ behaves approximately linearly with
temperature. This being the case,
it is now the bosonic term in Eq. 4 that determines the form of $\rho(T)$
everywhere, except at temperatures below $\sim t x^4$. 
Furthermore, assuming the $T=0$ c-axis superfluid
response to be also determined by the bosonic component fixes the 
parameter $t$: at low dopings 
$t= 2 ab \rho_c (0)/ ( x c^2)$, where $x/(abc)$ is the boson density,
$a$, $b$ and $c$ are the dimensions of the unit cell in the a-, b-, and
c-directions, and $\rho_c (0)$
the c-axis superfluid density (in Kelvins \cite{case}).  As an 
illustration, taking $x\approx ab/c^2 \approx 0.1 $
and the measured value  \cite{david}
of $\rho_c(0)/\rho (0) = 10^{-4} $, a satisfactory  
fit to experimental data \cite{broun} is achieved even 
by neglecting the quasiparticle term in Eq. 8 completely,
and by adjusting only $\rho_b(0)$  (Fig. 1).
In the overdoped regime, on the other hand, $\rho_b (T) \gg K(T)$,
and $\rho (T)$ should cross to the standard BCS result, 
as may have been already observed \cite{christos}.

For completeness, I also briefly describe 
the {\it non-superconducting phase} of $L$.
For $h<\sqrt{2\alpha}$, $\langle b_1 \rangle = \langle b_2 \rangle
=0$, and the bosons are in the incompressible phase. The integration
over the bosons then yields
\begin{equation}
L_\Phi \rightarrow \frac{K_\mu }{2} (v_\mu + A_\mu)^2 +
\frac{(\epsilon_{\mu\nu\rho} \partial_\nu v_\rho )^2 }{2m_b} +\frac{
 (\epsilon_{\mu\nu\rho} \partial_\nu a_\rho )^2}{2m_b},  
\end{equation}
to quadratic order, where $m_b ^2 \sim \alpha 
+O(\beta_{1,2} )$. At low energies one may still write 
$L=L_{sp}+L_{ch}$, but now with
$L_{sp}$ as the three dimensional quantum electrodynamics
($QED_3$)  \cite{franz}, \cite{igor2}.
Quasiparticles interact via long-range gauge interaction and
cease to be sharp excitations. Furthermore, fermions may
acquire a very small mass,  $\leq 10^{-4} m_b $ \cite{hands}, which would 
imply the antiferromagnetic ordering in the system. $L_{ch}$, on the other
hand, after the integration over $v_\mu$ becomes
\begin{equation}
L_{ch} \rightarrow 
\frac{(\epsilon_{\mu\nu\rho} \partial_\nu A_\rho )^2 }{2m_b}
 + \frac{i \epsilon_{\mu\nu\rho} \partial_\nu J_\rho
\epsilon_{\mu\alpha\beta} \partial_\alpha A_\beta }{K_\mu  m_b} +
\frac{J_\mu ^2}{2K_\mu }. 
\end{equation}
The first term implies that the system is an {\it insulator} with   a 
charge gap $\sim m_b$. The second term after a partial integration may
be rewritten as $\sim J_\mu \epsilon_{\mu\nu\rho}\partial_\nu B_\rho$,
where $B_\rho = \epsilon_{\rho\mu\nu}\partial_\mu A_\nu $, so 
a uniform magnetic field becomes completely decoupled in the insulator.
Since the lifetime of fermions
is inversely proportional to the fermion mass, and thus likely to 
be very long, thermal conductivity in the insulator
should essentially remain the same.
This could explain why the thermal conductivity of even a weakly insulating
YBCO apparently remains linear in temperature \cite{sutherland1}.

  It may be noteworthy that $L$ also has a {\it metastable} state
for $h>h_c$, with $\langle b_1 \rangle = 0$, and $2|\langle b_2 \rangle |^2
=(h^2 - h_c ^2 ) /(\beta_1 + \beta_2)$. In this state
$L_\Phi \rightarrow \rho_b (v_\mu +a_\mu )^2 /2 -ih n_b (v_0 +a_0 )$,
to quadratic order. Although $\rho_b \neq 0$, 
the full system is then actually metallic, since fermions acquire back 
their full electromagnetic charge and form four hole pockets with a small
Fermi surface, $\sim h$.

To summarize, the proposed effective theory predicts that
a weakly fluctuating d-wave BCS superconductor at large dopings crosses over
to the strongly phase-fluctuating  regime 
in the pseudogap, low-doping regime, and ultimately suffers a
transition into the insulator with possible antiferromagnetic order.
Superfluid density as a function of temperature
at low dopings becomes well approximated by the 
Bose condensate in the strongly anisotropic 3D Bose gas, and thus appears 
linear with an essentially  doping-independent slope
over most of the temperature range.

This work is supported by NSERC of Canada. I am grateful to D. Broun, 
D. Bonn, T. Lemberger, and J. Orenstein for providing me with their 
unpublished data, and to M. Case and P. Lee for
useful discussions.

\end{document}